\documentclass[allclo]{FBSart}
\usepackage{amsfonts}
\usepackage{amssymb}
\usepackage{graphicx}

\title{Three-Body Recombination in Ultracold Systems:
Prediction of Weakly-Bound Atomic Trimer Energies}

\author{Lauro Tomio\instnr{1}\thanks{\textit{E-mail:}
tomio@ift.unesp.br}, Victo S. Filho\instnr{1}, M. T. Yamashita\instnr{2},
A. Gammal\instnr{2}, \\ and T. Frederico\instnr{3}}
\instlist{Instituto de F\'\i sica Te\'orica, Universidade Estadual
Paulista, 01405-900, S\~{a}o Paulo, Brasil \and
Laborat\'orio do Acelerador Linear, Instituto de F\'\i sica
da USP, 05315-970, S\~ao Paulo, Brasil \and
Departamento de F\'\i sica, Instituto Tecnol\'ogico de
Aeron\'autica, CTA, 12228-900, S\~ao Jos\'e dos Campos, Brasil}
\runningauthor{L. Tomio et al. }
\runningtitle{Three-body recombination in ultracold systems: Prediction...}
\begin{document}
\maketitle
\begin{abstract}
The three-body recombination coefficient of a trapped ultracold atomic
system, together with the corresponding two-body scattering length $a$,
allow us to predict the energy $E_3$ of the shallow trimer bound state,
using a universal scaling function. The production of dimers in Bose-Einstein 
condensates, from three-body recombination processes, in the regime of short
magnetic pulses near a Feshbach resonance, is also studied in line with
the experimental observation.
\end{abstract}

\section{Introduction}
The interaction among neutral atoms is equivalent to a range zero in the
limit of $a\rightarrow \infty$. For identical bosonic atoms,
either the Efimov effect~\cite{efimov} or the 
Thomas collapse~\cite{thomas} can happen
when the ratio between $a$ and the range goes to infinity.
This is the scaling limit, where the three-boson properties can be
determined just by two scales: the dimer binding energy (or virtual
dimer energy) and the shallow trimer binding energy~\cite{amorim99}.

Recently, Bose-Einstein condensation (BEC) has been 
obtained with trapped ultra-cold atomic gases. Losses of atoms in BEC
can occur due to three-atom recombination processes,
which are measured in several cases:
$^{23}$Na$|F=1,m_F=-1\rangle$~\cite{stamper}, 
$^{87}$Rb$|F=1,m_F=-1\rangle$~\cite{burt,soding} and
$^{85}$Rb$|F=2,m_F=-2\rangle$~\cite{roberts,exp1}. 
The value of the three-body
recombination coefficient, together with the corresponding
two-body scattering length, allow us to predict the energy of the
shallow trimer bound state  for such systems in a trap, as we
describe in Sect. 2.

Another aspect related to the three-body recombination
process, supported by experimental observations, is a possible
enhancement due to quantum coherence (laser effect), which may
happen in a special situation of fast changes of $a$. In
Sect. 3, we discuss the dynamics of BEC related to this aspect
of the three-body recombination process.
Our conclusions are summarized in Sect. 4.

\section{Weakly bound atomic trimer energies}

It was shown by some of us, in ref.~\cite{rec03}, how to obtain
the trimer binding energies of a trapped atomic system from the
three-body recombination rate and the corresponding two-body
scattering length. A universal scaling function which gives the
dependence of the recombination parameter  as a function of the
weakly bound trimer energy ($E_3$) is calculated within a
renormalized scheme for three particles interacting through
pairwise Dirac-delta interaction~\cite{rec03}. The value of  $E_3$
are shown to be in the interval $1<m(a/\hbar)^2 E_3< 6.9$ for
large $a$. The contribution of a deep-bound state to the
prediction, in the case of $^{85}$Rb$|F=2,m_F=-2\rangle$, for a
particular trap, is shown to be relatively small. Also, using the
predicted energies of the triatomic molecules in ultra-cold traps
the sizes of the trimers $^{23}Na|1,-1\rangle$ and
$^{87}Rb|1,-1\rangle$ can be estimated. In this case, a universal
scaling function gives the mean-square radii as a function  of
$E_3$ and $a$. The root-mean-square distance between two atoms in
a triatomic molecule is estimated to be
0.8$\sqrt{\hbar^2/(m~E_3)}$, where $m$ is the atomic
mass~\cite{raio02}.

The rate of three free bosons to recombine, forming a dimer and one
remaining particle, is given in the limit of zero energy, by the
recombination coefficient\cite{nielsen,fedi}
\begin{eqnarray}
K_3=\kappa\frac{\hbar}{m}a^4 ,
\label{k3}
\end{eqnarray}
where $\kappa$ is a dimensionless parameter.
When $a>0$, the recombination parameter $\kappa$
oscillates between zero and a maximum value, which is a function
of $a$, as shown in refs.~\cite{nierec} ( $\kappa\le 68.4$ ),
\cite{esryrec} ($\kappa\le 65$) and \cite{bedarec} ($\kappa\le
67.9$).

By using a general scaling procedure, it is shown in ref.~\cite{rec03}
that one can express the functional dependence of $\kappa$ in terms of
the two and three-body energies,
$\kappa\equiv\kappa\left( \sqrt{{E_2}/{E_3}} \right)$,
considering that, for large scattering lengths one has
$1/a=\sqrt{mE_2/\hbar^2}$.
The scaling form of $\kappa$, with explicit
$\sqrt{E_2/E_3}$ dependence, is 
\begin{eqnarray}
\kappa = \kappa_{\rm max}\sin^2\left(-1.01\ln\sqrt{\frac{E_2}{E_3}} +
\Delta\left(\sqrt{\frac{E_2}{E_3}}\right)\right) \ ,
\label{als}
\end{eqnarray}
where
$\Delta\left(\sqrt{E_2/E_3}\right)=\delta -
1.01\ln\left(\sqrt{{mE_3/\hbar^2}}\right)$ and
$\delta$ depends on the interaction at short distances.

Next, considering the calculation of the scaling function, with
the subtracted form of the Faddeev equations~\cite{ren},
and considering the relation between the transition matrix
elements and the three-body recombination coefficient at zero
energy, it was derived the functional dependence of $\kappa$ in
terms of the ratio $E_2/E_3$ (See Fig. 1 of ref.~\cite{rec03}).
The scaling limit is well approached in the calculations, with the
maximum $\kappa$ occurring at the threshold ($E_3=E_2$) and when
$(E_3/E_2)^{\frac12}=$0.38~\cite{amorim99}, such that between
two consecutive maximum of $\kappa$ we have $ 1<m(a/\hbar)^2 E_3<
6.9$.

Using the experimental values $\kappa_{expt}$, for a few
atomic species $^AZ|F,m_F\rangle$, one can obtain two possible values of
the weakly bound triatomic molecular states that are consistent with the
universal scaling plot that was derived.
The predicted trimer binding energies are given in
Table 1, in miliKelvins, in respect to the threshold, $S_3\equiv
(E_3-E_2)$ and $S_3^{\prime}\equiv(E'_3-E_2)$,
considering the central values of the experimental dimensionless
recombination parameters.

\begin{table}
\caption[dummy0]
{For the atomic species $^AZ|F,m_F\rangle$, it is shown in the 4th and
5th columns the predicted trimer binding energies, in respect to the
threshold. $\kappa_{expt}$ and $E_2$ are respectively given in the 2nd
and 3rd column. For $^{87}$Rb$|1,-1\rangle$, the
recombination process was obtained in ref.~\cite{burt} for noncondensed
($^*$) and condensed ($^\dagger$) trapped atoms.}
\vspace{-0.2cm}
\begin{center}
\begin{tabular}{ccccc}
\hline\hline
$^AZ|F,m_F\rangle$ & $\kappa_{expt}$ &$E_2$ (mK) &
$S_3$(mK) & $S_3^{\prime}$(mK)\\
\hline
$^{23}$Na$|1,-1\rangle$ & 42$\pm$12~\cite{stamper}
& 2.85 & 4.9 & 0.21\\
$^{87}$Rb$|1,-1\rangle$ & 52$\pm$22$^*$~\cite{burt}
& 0.17 & 0.39 & 0.005\\
$^{87}$Rb$|1,-1\rangle$ & 41$\pm$17$^\dagger$~\cite{burt}
&0.17 & 0.30 & 0.013\\
$^{87}$Rb$|2,2\rangle$ & 130$\pm$36~\cite{soding}
&0.17 & -    & - \\
$^{85}$Rb$|2,-2\rangle$
& 7.84$\pm$3.4~\cite{roberts,exp1}
&1.3$\times 10^{-4}$
&1.14$\times 10^{-4}$ & 3.8$\times 10^{-5}$ \\
\hline\hline
\end{tabular}
\end{center}
\end{table}

\section{Coherent dimer formation}

One should note that the three-body recombination coefficient was
derived in the previous section considering very dilute systems
and considering that the two-body interaction is kept fixed.
Another interesting possibility may happen in a special situation
when we have fast changes of the two-body scattering length $a$.
In this case, the three-body recombination process may be
enhanced due to quantum coherence. The remaining atoms and dimers
are produced in a single state and, due to the symmetrization of
the full wave-function, the recombination process which happens in
this background is enhanced (laser effect).

Supporting this possibility, it was observed~\cite{exp1} a burst
of relatively hot $^{85}$Rb atoms with temperature of about 150
nK, when a fast magnetic pulse near a Feshbach resonance increases
the scattering length up to $4000 a_0$. The maximum value of $a$
gives a lower bound for the temperature of the remaining atoms
about 75 nK consistent with the experiment. In line with the
experimental observation, we make an attempt to describe the 
remaining number of condensed atoms after the pulse, within  a 
mean field description. For that, we consider that: 
i) shallow dimers are formed
(the temperature of the atoms in the burst is consistent with
weakly bound dimers and not deeply bound dimers); 
ii) the recombination process shows a coherence effect (the burst supports
it); and 
iii) a fitted magnitude of the three-body recombination
parameter depends on the rise time ($t_r$) of the
scattering length [Also using that $K_3$ is proportional to $a(t)^4$].

One has to go beyond the mean-field model to implement the
coherence effect in a defined theory. The phenomena should be
described in a coupled model, with the condensed atoms, the
remaining atoms and the dimers. In this respect, it was
suggested recently, in a theoretical microscopic approach~\cite{Borca}, 
that in a collision of two atomic condensates producing two molecular 
condensates in counterpropagating momentum eigenstates, 
it is possible to appear an atom-molecule
laser fed by stimulated three-body recombination processes,
where atoms and molecules produced in the same state enhance the
three-body recombination rate in respect to the vacuum values.

In an  effective way, i.e., using the mean-field equation, the
coherence effect (laser effect) will appear as the increase of the
magnitude of the recombination parameter in respect to the vacuum
value. As the process becomes faster, the loss of coherence of the
atoms and dimers is reduced; and, as suggested by our results, the
increase in the three-body recombination rate is even 
higher~\cite{dimerlas}.

For describing the dynamics of BEC as nonconservative systems,
we can follow the extended mean field formalism developed in
ref.~\cite{vsf}, in which one considers an extended approach of
the Gross-Pitaevskii formalism, by including
losses of the system by three-body recombination. Here, we
should also include the time variation of the relevant physical
parameters.

Near a Feshbach resonance, the scattering length $a$ has
been observed to vary as a function of magnetic field B, according
to theoretical prediction \cite{Moerdijk}, as
\begin{equation}
a=a_{b} \times \left(1-\frac{\Delta}{B-B_r}\right)\,\,\,,
\label{scat}
\end{equation}
where $a_b$ is the background scattering length, $B_r$ is the
resonance magnetic field and $\Delta\equiv (B_0-B_r)$ is the
resonance width, where $B_0$ is the value of $B$ at $a=0$. For the
case of $^{85}$Rb, one has $\Delta \cong $ 11.0 G, $B_{r} \cong $
154.9 G and $a_b \cong $ -450 $a_0$. We consider in our
calculations the same experimental parameters used in
ref.~\cite{exp1}: initial field B$_0 \cong $ 166 G (harmonic
oscillator state), applied to an initial sample of N$_0$=16500
condensed atoms; spherical symmetry, with mean geometric frequency
${\omega} \cong $ 2$\pi\times$ 12.77 Hz, for simulating the
cylindrical geometry of JILA ($\omega_r = $2$\pi\times$ 17.5 Hz
and $\omega_z = $2$\pi\times$ 6.8 Hz).

For the above conditions, the remaining number of atoms in BEC was
obtained and compared with JILA experimental data. The results are
given in Table 2. We note that, in order to obtain the best fit of
the experimental data, we need to adjust $\kappa$ to
values much higher than the vacuum values given by 
Eq.~(\ref{k3}).
For instance, if we consider the rise time of the magnetic field
pulses lower than 100 $\mu$s, we have 2000 $\lesssim \kappa
\lesssim$ 1000. However, $\kappa$ decreases while $t_r$ increases,
such that $\kappa\sim$ 100 for $t_r$ of the order of 250 $\mu$s.
In this case, slowing the rise time, we note that $\kappa$
approaches the predicted interval 0 $ \le \kappa \lesssim$ 68.

\begin{table}
\caption{Numerical values of the three-body recombination coefficient
$\kappa$ as function of the rise time $t_r$ of the magnetic field pulses
applied to the $^{85}$Rb BEC.}
\begin{center}
\begin{tabular}{c|rrrrrr}
\hline\hline
$t_r$ ($\mu$s)& 12.5 & 25.3 & 75.8 & 151.6 & 202.1 & 252.6 \\
\hline
$\kappa$      & 1800 & 1700 & 1600 &  500  & 200   & 100 \\
\hline\hline
\end{tabular}
\end{center}
\end{table}

The phenomenon of enhancement in the magnitude of the three-body
recombination coefficient gives an indication that may be a
laser-like effect is occurring in the experiment of
ref.~\cite{exp1}.  Our picture suggests that for very short
time scales (time ranges varying from 0 up to 100 $\mu$s) the
three-body recombination processes produce shallow dimers
and a third atom (burst effect) in many-boson states. The
coherent production of shallow dimers occurs  for very short
rise times and disappears for longer ones, producing the strange
effect reported in ref.~\cite{exp1}, of a decreasing dissipation
from the condensate for long rise times. This effect is brought to
the mean-field calculation by the enhanced values of $\kappa$ as a
function of $t_r$, which  can be approximately  given by the
parameterization
\begin{eqnarray}
\kappa(t_r)&=&2300\exp{(-0.01\times\omega t_r)}\;.
\label{adj1}
\end{eqnarray}
Such parametrization is convenient to study the remaining 
number of atoms in $^{85}$Rb BEC as a function of the scaled 
rise time of the applied magnetic pulse, for fixed hold times, 
with the parameters considered in ref.~\cite{exp1} 
(where the magnetic field during the hold time is 
$B_h=$ 156.7 G). The results of our model calculation 
describes the observed behavior of the remaining
number of atoms (See Fig. 3 of ref.~\cite{dimerlas}).

\section{Conclusions}

In the present communication, we report recent results that we
have obtained by considering the three-body recombination 
processes in ultracold or condensed atomic gases. From the
experimental values of the recombination coefficient together with
the corresponding two-body scattering length, for atomic systems
in a trap, we are able to predict energies of the shallow trimer
bound state for a few atomic systems. To obtain the scaling
function corresponding to the dependence of the dimensionless
recombination parameter with the shallow trimer binding energy, we
use a zero range model which is valid for large scattering
lengths. We point out that, at least one experimental result for
$^{87}$Rb$|2,2\rangle$ is outside the interval of theoretical
vacuum values of $\kappa$, which may indicate that some particular
physics is present in that experiment.

We also observe that, when considering the special situation of
fast variations of the two-body scattering length, near a Feshbach
resonance, the experimental results can be put consistent with a
mean field description, if one allows a strong enhancement of the
three-body recombination process motivated by a possible quantum
coherence phenomena in the production of dimers (laser effect).
However, certainly one has to go beyond the mean-field description
with only condensed atoms, to be able to implement the coherence
effect in a defined way. Such a mean-field theory should describe
simultaneously the dynamics of the condensed atoms, the remaining
atoms and the dimers, which is within the scope of a future
investigation.

\begin{acknowledge}
This work was supported by Funda\c c\~ao de Amparo \`a Pesquisa do
Estado de S\~ao Paulo and Conselho Nacional de Desenvolvimento
Cient\'\i fico e Tecnol\'ogico.
\end{acknowledge}

\end{document}